\documentclass{elsart}
\usepackage[dvips]{graphicx}
\newcommand{\krzb}{\ensuremath{\overline{K}^{*0}}}
\newcommand{\krzmndk}{\ensuremath{D^+ \rightarrow \krzb \mu^+ \nu}}
\newcommand{\krzlndk}{\ensuremath{D^+ \rightarrow \krzb \ell^+ \nu_\ell}}
\newcommand{\philndk}{\ensuremath{D_s^+ \rightarrow \phi\; \ell^+ \nu_\ell}}
\newcommand{\phimndk}{\ensuremath{D_s^+ \rightarrow \phi\; \mu^+ \nu }}

\newcommand{\kpimndk}{\ensuremath{D^+ \rightarrow K^- \pi^+ \mu^+ \nu }}

\newcommand{\gevcsq}{\ensuremath{\textrm{GeV}/c^2}}

\newcommand{\mkpi}{\ensuremath{m_{K\pi}}}

\newcommand{\krzlnbrb}{\ensuremath{{\Gamma(\krzlndk{}) \over \Gamma(\kpipi{})} }}
\newcommand{\krzmnbrb}{\ensuremath{{\Gamma(\krzmndk{}) \over \Gamma(\kpipi{})} }}
\newcommand{\phimnbrb}{\ensuremath{{\Gamma(\phimndk{}) \over \Gamma(\phipi{})} }}
\newcommand{\philnbrb}{\ensuremath{{\Gamma(\philndk{}) \over \Gamma(\phipi{})} }}
\newcommand{\phimnbr}{\ensuremath{{\Gamma(\phimndk{}) / \Gamma(\phipi{})} }}
\newcommand{\kpipi}{\ensuremath{D^+ \rightarrow K^- \pi^+ \pi^+ }}
\newcommand{\phipi}{\ensuremath{D_s^+ \rightarrow \phi\; \pi^+ }}
\newcommand{\krzresult}{\ensuremath{ 0.602 ~\pm~0.010~{\rm(stat)}~\pm~0.021~{\rm (sys)}}}
\newcommand{\krzresultstat}{\ensuremath{ 0.602 ~\pm~0.010~}}

\newcommand{\phiresult}{\ensuremath{0.54~\pm~0.033~{\rm (stat)}~\pm~0.048~{\rm (sys)}}}
\newcommand{\phiresultstat}{\ensuremath{0.54~\pm~0.033~}}
\newcommand{\mysection}[1]{\section{#1}}
\newcounter{saveeqn}%

\begin{document}
\begin{frontmatter}
\title{New Measurements of the \krzmnbrb{} and \phimnbrb{} Branching Ratios}
The FOCUS Collaboration%
\footnote{See \textrm{http://www-focus.fnal.gov/authors.html} for
additional author information.}
\author[ucd]{J.~M.~Link}
\author[ucd]{M.~Reyes}
\author[ucd]{P.~M.~Yager}
\author[cbpf]{J.~C.~Anjos}
\author[cbpf]{I.~Bediaga}
\author[cbpf]{C.~G\"obel}
\author[cbpf]{J.~Magnin}
\author[cbpf]{A.~Massafferri}
\author[cbpf]{J.~M.~de~Miranda}
\author[cbpf]{I.~M.~Pepe}
% ...

\address[ucd]{University of California, Davis, CA 95616}
\address[cbpf]{Centro Brasileiro de Pesquisas F\'isicas, Rio de Janeiro, RJ, Brasil}
\address[cinv]{CINVESTAV, 07000 M\'exico City, DF, Mexico}
\address[cu]{University of Colorado, Boulder, CO 80309}
\address[fnal]{Fermi National Accelerator Laboratory, Batavia, IL 60510}
\address[fras]{Laboratori Nazionali di Frascati dell'INFN, Frascati, Italy I-00044}
\address[ui]{University of Illinois, Urbana-Champaign, IL 61801}
\address[iu]{Indiana University, Bloomington, IN 47405}
\address[korea]{Korea University, Seoul, Korea 136-701}
\address[kp]{Kyungpook National University, Taegu, Korea 702-701}
\address[milan]{INFN and University of Milano, Milano, Italy}
\address[nc]{University of North Carolina, Asheville, NC 28804}
\address[pavia]{Dipartimento di Fisica Nucleare e Teorica and INFN, Pavia, Italy}
\address[pr]{University of Puerto Rico, Mayaguez, PR 00681}
\address[sc]{University of South Carolina, Columbia, SC 29208}
\address[ut]{University of Tennessee, Knoxville, TN 37996}
\address[vu]{Vanderbilt University, Nashville, TN 37235}
\address[wisc]{University of Wisconsin, Madison, WI 53706}

\nobreak
\begin{abstract}
Using a large sample of charm semileptonic decays collected by the
FOCUS photoproduction experiment at Fermilab, we present new
measurements of two semileptonic branching ratios.  We obtain values
of \krzmnbrb{}~=~\krzresult{} and
\phimnbrb{}~=~\phiresult{}. Our \krzmndk{} result includes the effects of the 
s-wave interference discussed in Reference \cite{anomaly}.
\end{abstract}
\end{frontmatter}
\newpage
% No page number printed for this page
%\tableofcontents    % contents are based on the sections, subsections, etc.
%\listoffigures     % based on the figures
\newpage

\mysection{Introduction}

This paper describes new measurements of charm semileptonic decay
rates into the \krzmndk{} and \phimndk{} final states relative to the
hadronic decay rates \kpipi{} and \phipi{} respectively.  The
\krzmndk{} width sets the scale of the $A_1(0)$ form factor describing
these decays \cite{KS} and therefore tests Lattice Gauge and quark
model calculations. Such tests of the overall form factor scale are
particularly relevant to the determination of CKM matrix element
ratios such as $|V_{cd}/V_{cs}|^2$ which relates to (for example)
$\Gamma (D^+ \rightarrow \rho \mu^+ \nu)/ \Gamma (D^+ \rightarrow \bar
K^* \mu^+ \nu)$.  The \phimndk{} process has been frequently used to
estimate the $D_s^+$ branching ratios via assumptions about
$\Gamma(\philndk{})$ since no high statistics double tag deduced
branching fractions for the $D_s^+$ presently exist. Frequently the
$D_s^+$ yields are established by counting \phipi{} decays making
a good measurement of \phimnbr{} important.

Recently the CLEO Collaboration obtained a new measurement of
\krzlnbrb{} that is somewhat higher than previous measurements. 
They state in Reference \cite{cleo}, that their new value implies an
increase in the ratio ${\Gamma(D \rightarrow \overline{K}^* \ell^+
\nu) \over \Gamma(D \rightarrow K^- \ell^+ \nu)}$ bringing it more in
line with early quark model estimates and in considerable discrepancy
with the ISGW2 model \cite{isgw2}. This Letter discusses a more
precise determination of the \krzmnbrb{} from FOCUS and provides a
first measurement of this ratio that includes the effects of the
interfering s-wave described in our recent paper \cite{anomaly}.  The
interference of this s-wave amplitude with the dominant \krzmndk{} 
contribution to \kpimndk{} significantly distorts the angular decay
distributions thus affecting the reconstruction efficiency for
this state. Throughout this paper, unless explicitly stated otherwise,
the charge conjugate is also implied when a decay mode of a specific
charge is stated.

\mysection{Experimental and analysis details}

The data for this paper were collected in the Wideband photoproduction
experiment FOCUS during the Fermilab 1996--1997 fixed-target run. In
FOCUS, a forward multi-particle spectrometer is used to measure the
interactions of high energy photons on a segmented BeO target. The
FOCUS detector is a large aperture, fixed-target spectrometer with
excellent vertexing and particle identification. Most of the FOCUS
experiment and analysis techniques have been described
previously~\cite{ycp}. The FOCUS muon system and typical
analysis cuts are described in References \cite{nim} and \cite{anomaly}.

To isolate the \kpimndk{} topology, we required
that candidate muon, pion, and kaon tracks appeared in a secondary
vertex with a confidence level exceeding 1\%.  The muon track, when
extrapolated to the shielded arrays of the inner muon system located
downstream of both magnets, was required to match muon
hits with a confidence level exceeding 5\%. The kaon was required to
have a \v Cerenkov light pattern more consistent with that for a kaon
than that for a pion by 1 unit of log likelihood, while the pion
track was required to have a light pattern favoring the pion
hypothesis over that for the kaon by 1 unit~\cite{CNIM}.

To further reduce muon misidentification, a muon candidate was allowed
to have no or one missing hit in the 6 planes comprising our inner
muon system. In order to suppress muons from pions and kaons decaying
in our spectrometer we required that muon candidates have an
energy exceeding 10 GeV.  Non-charm and random combinatoric
backgrounds were reduced by requiring both a detachment between the
vertex containing the $K^-\pi^+\mu^+$ and the primary production
vertex of 10 standard deviations and a reconstructed energy between 40
and 180 GeV.  Background from $D^+ \rightarrow K^- \pi^+ \pi^+$, where
a pion is misidentified as a muon, was reduced by requiring that
the visible mass $m_{K\pi\mu} < 1.8~\gevcsq{}$. In order to suppress
background from $D^{*+} \rightarrow D^0 \pi^+ \rightarrow (K^- \mu^+
\nu) \pi^+$ we required $M(K^- \mu^+ \nu \pi^+) - M(K^- \mu^+ \nu) >
0.18~\gevcsq $. The momentum of the undetected neutrino was estimated
from the $D^+$ line-of-flight as discussed in Reference
\cite{anomaly}.

In addition to these cuts, that we will call ``baseline'' cuts, we
imposed the following two cuts on the sample that we use to quote the
branching ratio.  To suppress possible backgrounds from higher
multiplicity charm decay, we isolated the $K\pi\mu$ vertex from other
tracks in the event (not including tracks in the primary vertex) by
requiring that the maximum confidence level for another track to form
a vertex with the candidate be less than 0.1\%.  To suppress background
from the re-interaction of particles in the target region which can mimic
a decay vertex, we required that the charm
secondary vertex was located at least three measurement standard
deviations outside of all solid material including our target and
target microstrip system.  We will call this the ``out-of-material''
cut.

Apart from the muon cuts and the cut on the $M(K^- \mu^+ \nu \pi^+) -
M(K^- \mu^+ \nu)$ mass difference, these same cuts were applied to our
\kpipi{} sample.  We required one of the two pions in the \kpipi{}
final state to be in the inner muon system to better match the
angular region required for the muon in \kpimndk{} events.

We turn next to a discussion of the cuts used in the \phimndk{}
analysis. Most of the cuts used for the \kpimndk{} and \kpipi{}
selection were used to select our $\phi \mu^+ \nu$ and $\phi \pi^+$
samples. Because the $D_s^+$ lifetime is shorter than the $D^+$, we
reduced our cut on the primary-secondary vertex detachment to greater
than 5 standard deviations.  To further reduce non-charm background we
required that our primary vertex consisted of at least two charged
tracks.  To further reduce muon contamination to the \phimndk{} state
due to decays of pions and kaons in flight, we required that the 
confidence level that a muon track had a consistent
trajectory through the two magnets comprising the FOCUS magnetic
spectrometer exceeded 5\%.

\mysection{Analysis of the \krzmndk{} final state}

\begin{figure}[tbph!]
 \begin{center}
\includegraphics[height=1.9in]{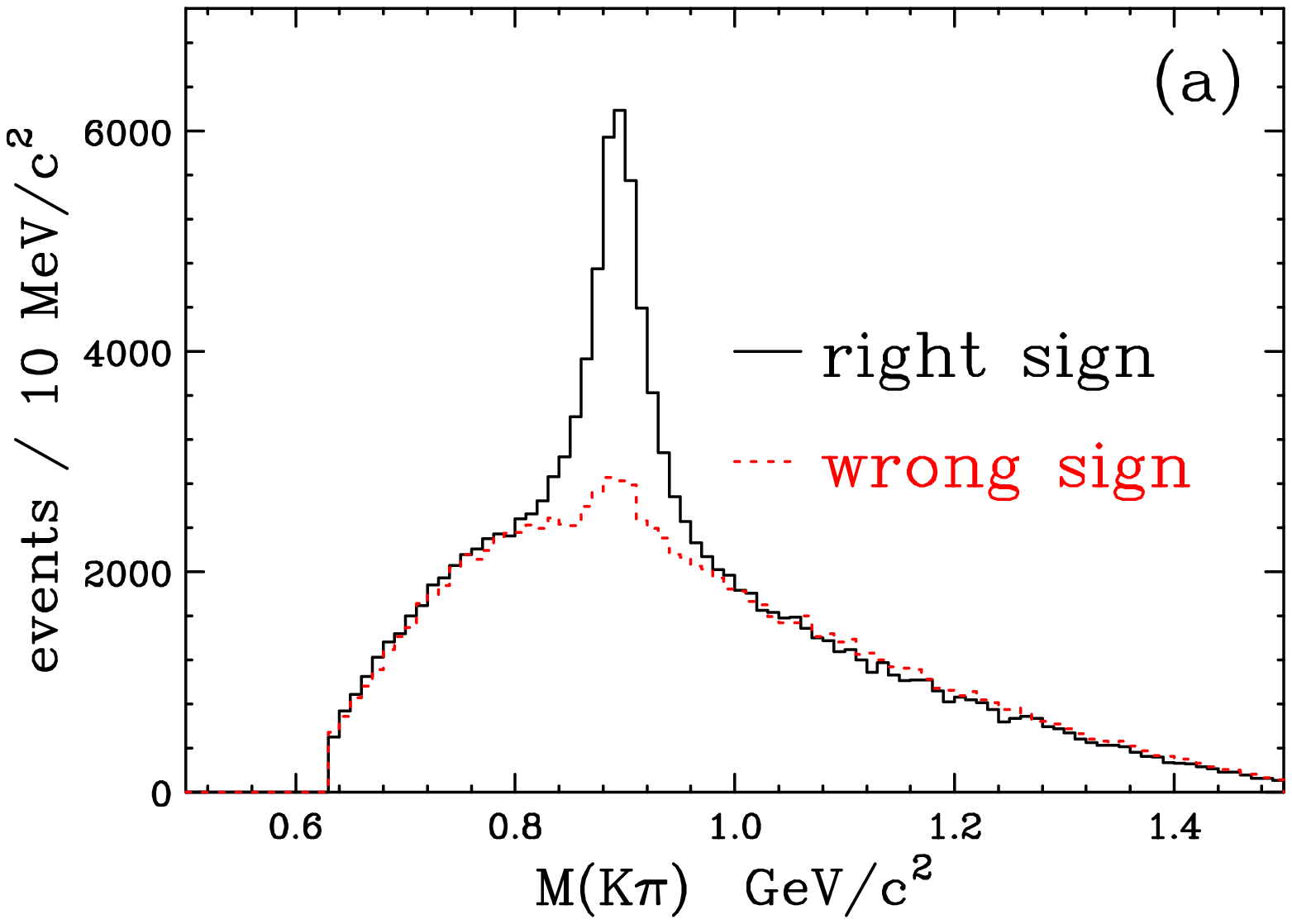}
\includegraphics[height=1.9in]{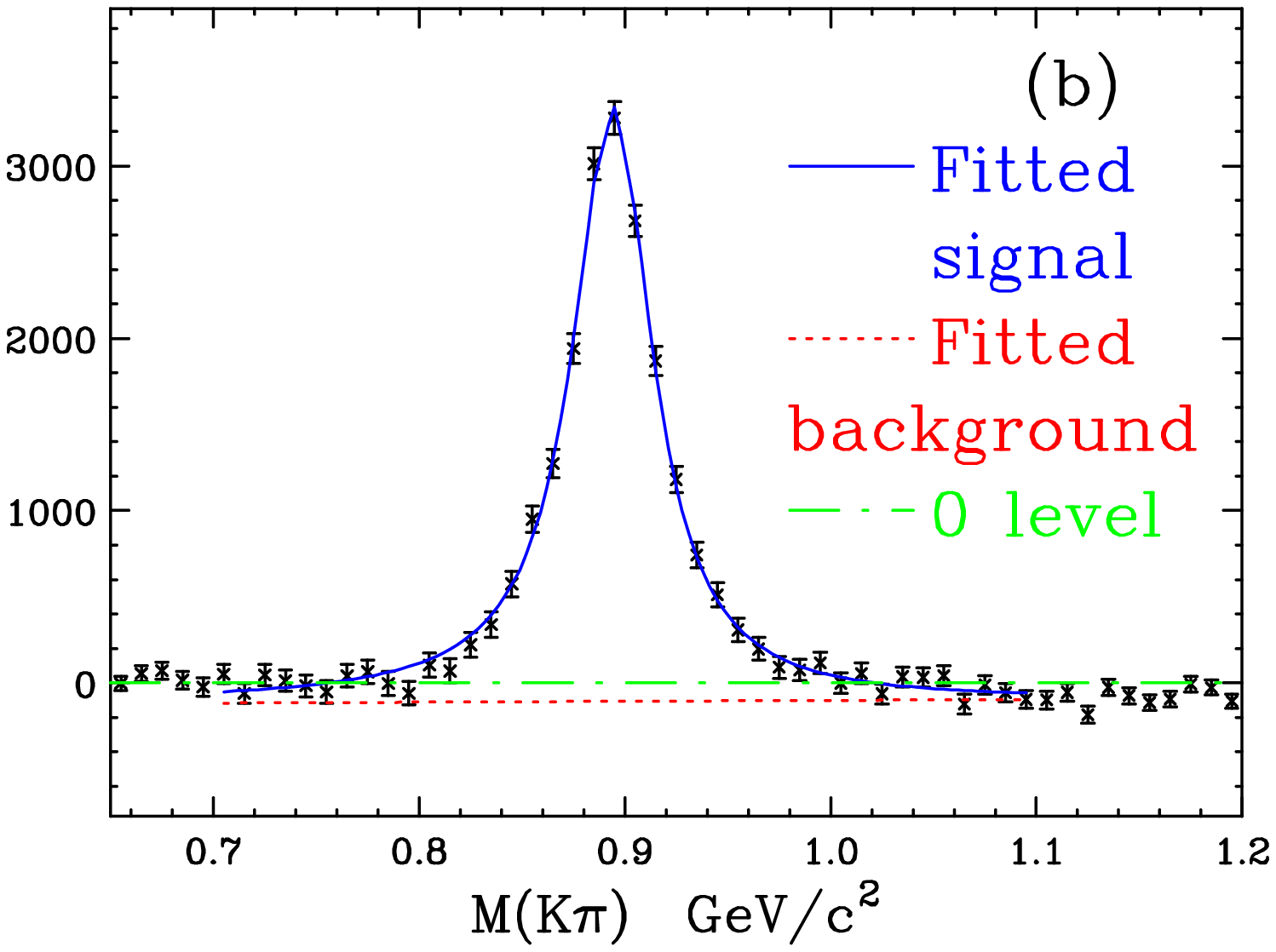}
\includegraphics[height=1.9in]{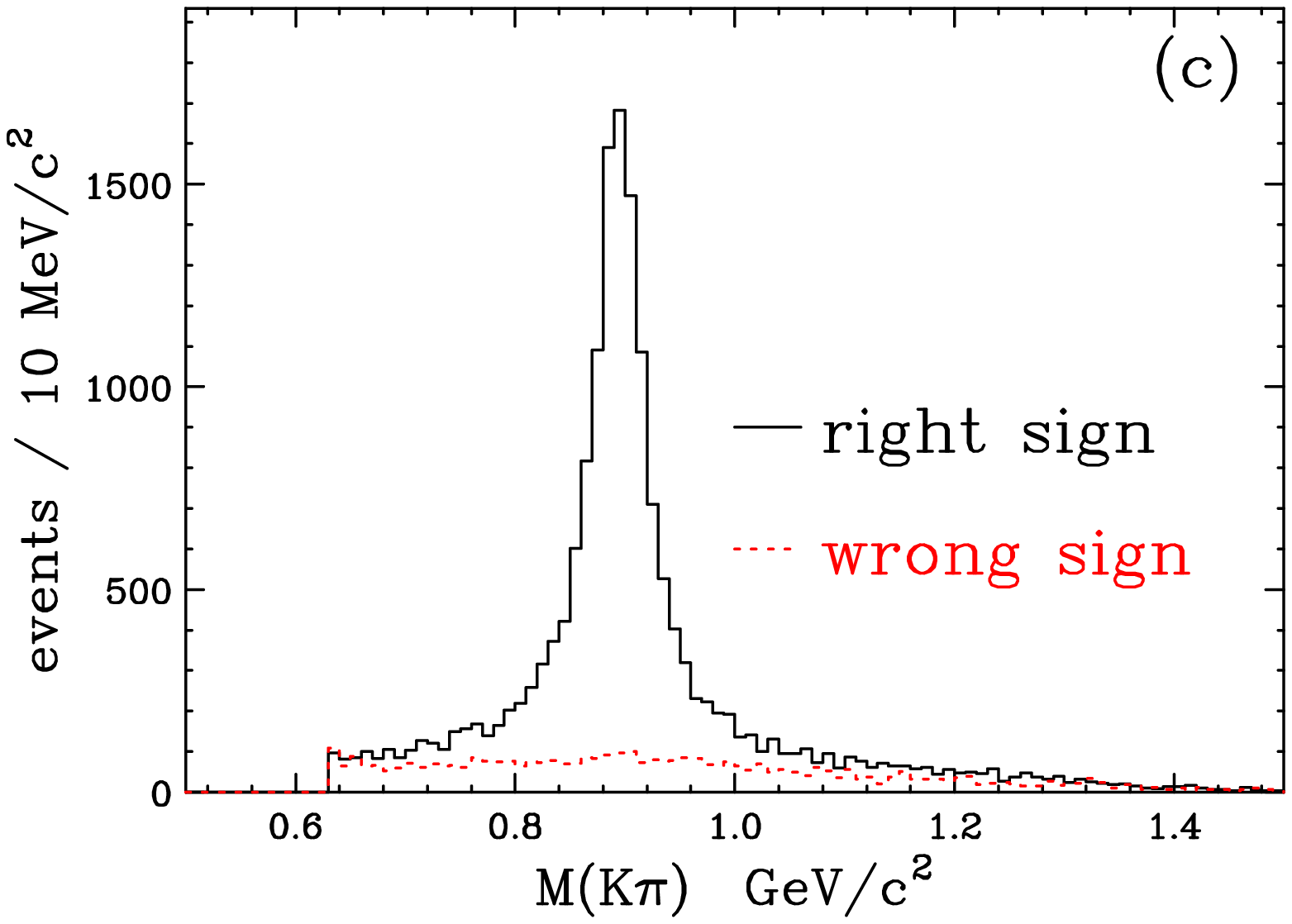}
\includegraphics[height=1.9in]{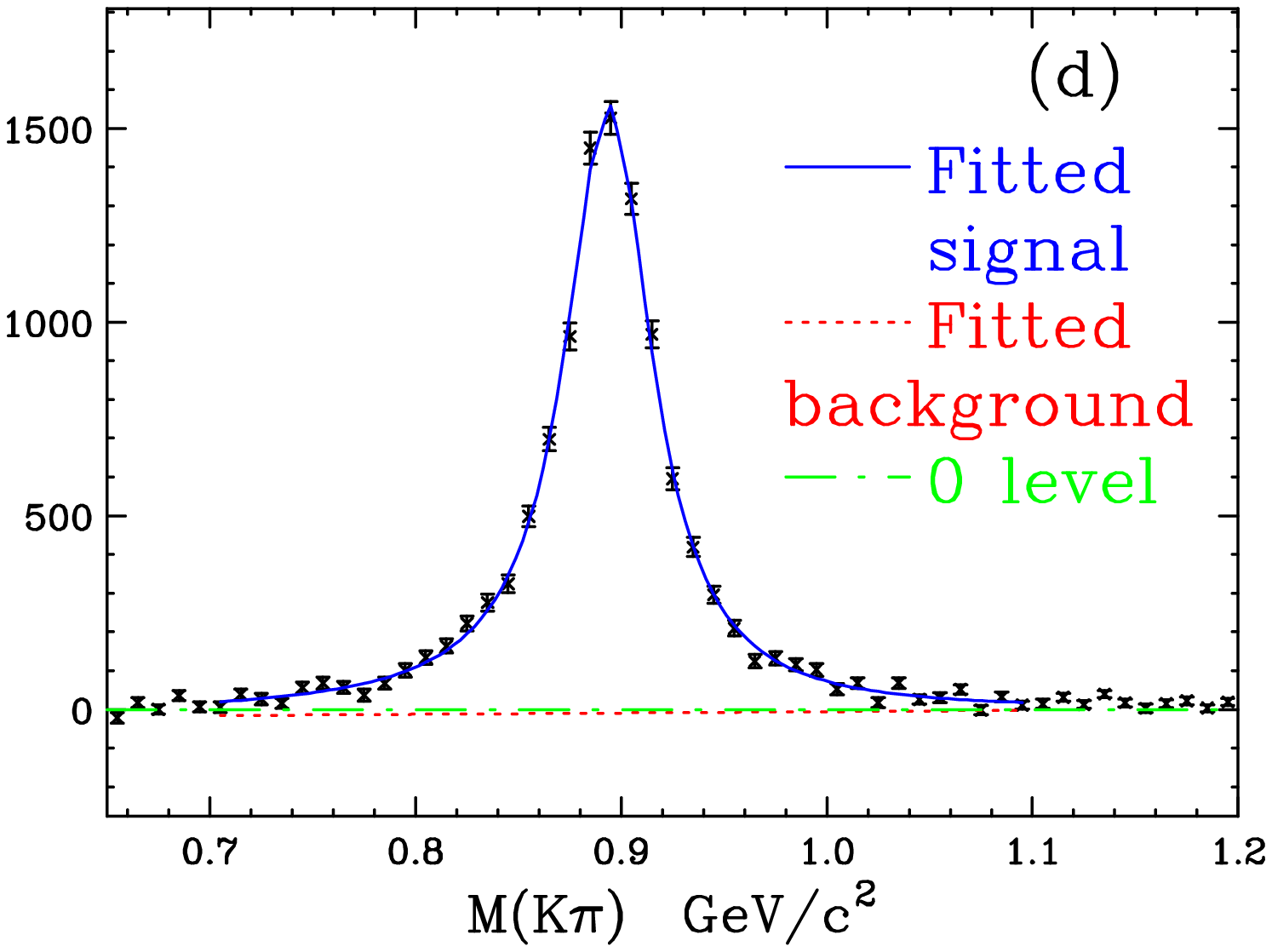}
\caption{We show two $D^+ \rightarrow K^* \mu\nu$ signals and fits (a)
The $K^- \pi^+$ mass spectra subjected to the baseline cuts. The right-sign
signal is plotted in a solid line and wrong-sign signal is plotted in a dotted
line.  (b) Fits
to the subtracted baseline signal (described in the text). Both the wrong-sign,
and charm
backgrounds are subtracted prior to fitting.  The fit, described in
the text, consists of a Breit-Wigner plus a 1st order polynomial.  The
\krzmndk{} yield obtained from this fit consists of 23\,726 events.
(c) The $K^- \pi^+$ mass spectra with an additional out-of-material
cut and secondary isolation cut. The right-sign signal is plotted in a solid
line and
wrong-sign signal is plotted in a dotted line.  (d) Fits to the baseline signal
with
additional out-of-material and secondary isolation cut. The \krzmndk{}
yield obtained from this fit consists of 11\,698 events.
\label{signal}}
\end{center}
\end{figure}

Figure~\ref{signal} shows the $K^- \pi^+$ mass distributions and fits
of the signal we obtained using two selections of the cuts described
above.  A very strong \krzb{}(896) signal is present for both
samples. To assess the level of non-charm backgrounds, we plot the
``right-sign'' (where the kaon and muon have the opposite charge) and
``wrong-sign'' $K^- \pi^+$ mass distributions separately in Figure
\ref{signal} (a) and (c) for the sample with the baseline cuts and the
sample with baseline, out-of-material, and isolation cuts respectively.
We attribute the dramatic decrease in the wrong sign component in
Figure \ref{signal} (c) compared to Figure \ref{signal} (a) to a
reduction in non-charm backgrounds through the use of the
out-of-material cut.  

The fits to the subtracted $K^- \pi^+$ mass distribution shown Figure
\ref{signal} (b) and (d) are used to estimate the yield of \krzmndk{}
events.  Two subtractions are applied.  We first subtract the
distributions of wrong-sign from right-sign events as a means of
subtracting non-charm backgrounds that are nearly charge symmetric.
We then subtract the anticipated right-sign excess predicted by our
Monte Carlo simulation which simulates all known charm decay
backgrounds.  The Monte Carlo distribution used in this subtraction 
is scaled by the ratio of the fitted yield of \kpipi{} events observed
in the data to that observed in the Monte Carlo.   This second
subtraction both simplifies the shape of the non-\krzb{} component of the 
observed $K^- \pi^+$ mass spectrum and corrects for the presence
of \krzb{}'s from charm sources other than \krzmndk{} such as hadronic
charm decays where one of the secondaries is misidentified as a muon.

The fits overlayed on Figure \ref{signal} (b) and (d) are relativistic
Breit-Wigner line shapes plus a 1st order polynomial of the form
$\alpha (\mkpi{} - m_o ) + \beta$ where $m_o$ is the pole mass of
the \krzb{}.  The range of the fit is $0.7 < \mkpi{} < 1.1
~\gevcsq{}$.  The fit parameters were the resonant Breit-Wigner mass,
yield, and width and the $\alpha$ and $\beta$ background parameters.

We found the subtracted data better fit a constant width Breit-Wigner
with a mass and width consistent with the known\cite{PDG} parameters
of the \krzb{} rather than the conventional p-wave
form. We performed a variant of the subtracted
$K^- \pi^+$ fit where the width was allowed to vary with mass
according to $\Gamma(\mkpi{}) = \Gamma_o (p/p_o)^N$ where $p$ is the
kaon momentum in the kaon-pion rest frame at a given \mkpi{}, $p_o$ is
value of this momentum when the kaon-pion mass equals the resonant
mass $m_o$ and the power $N$ was a fit parameter.  We found that the
data was very consistent with the power $N=0$ and inconsistent (at the
20 $\sigma$ level) with $N = 3$ -- the value expected for a p-wave
Breit Wigner. We plan to present, in a subsequent publication, a
detailed analysis of the \mkpi{} line shape in \krzmndk{} decays
including the effects of interference from the s-wave amplitude
\cite{anomaly} and possible other sources, efficiency variation, and
mass dependent form factor and barrier corrections. 

Figure \ref{signal} (b) shows the presence of a significant negative offset
($\beta$) that is nearly absent in \ref{signal} (d) which we attribute
to a broad structure excess of wrong-sign events in our non-charm
background.
\begin{figure}[tbph!]
 \begin{center}
\includegraphics[height=1.9in]{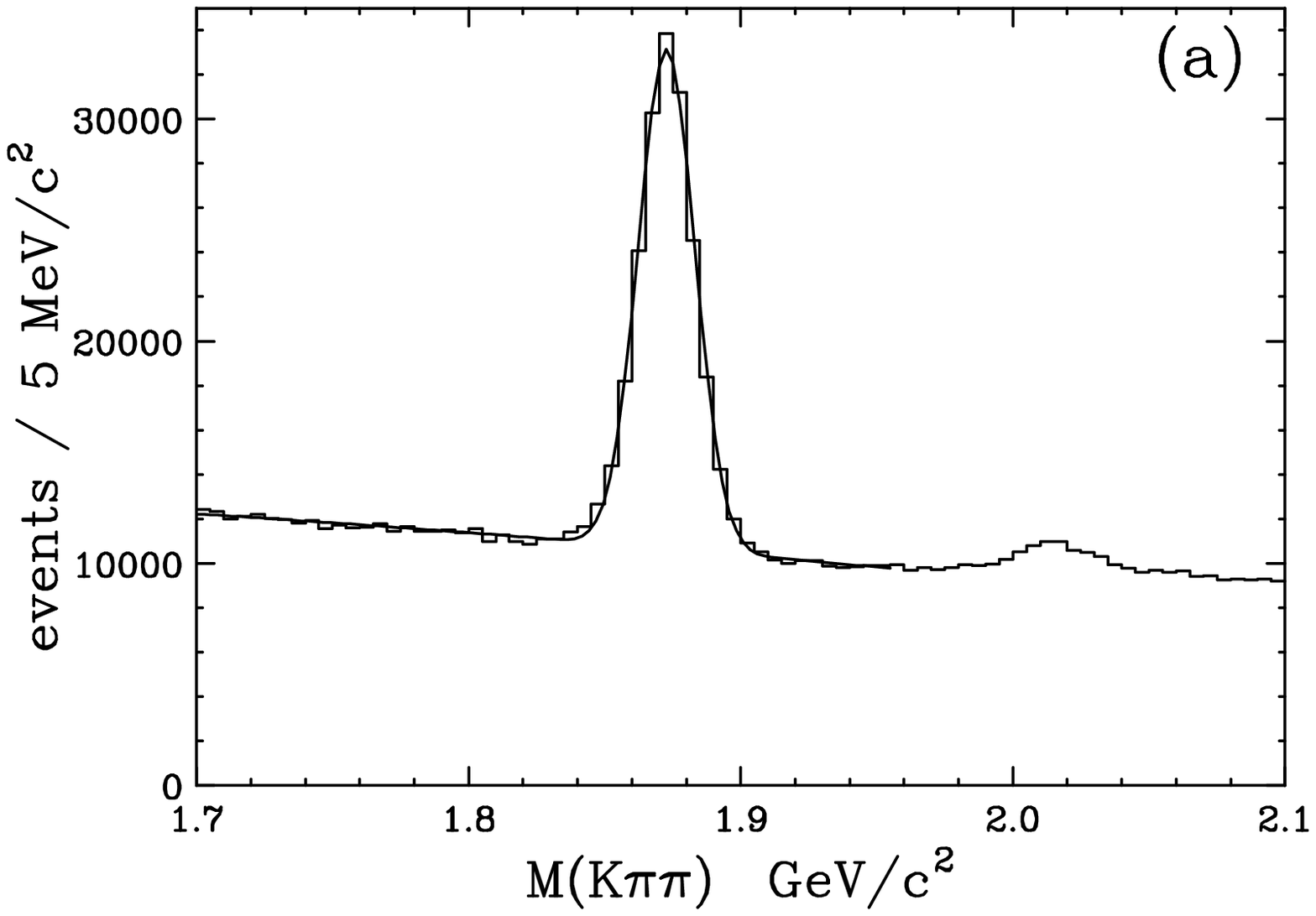}
\includegraphics[height=1.9in]{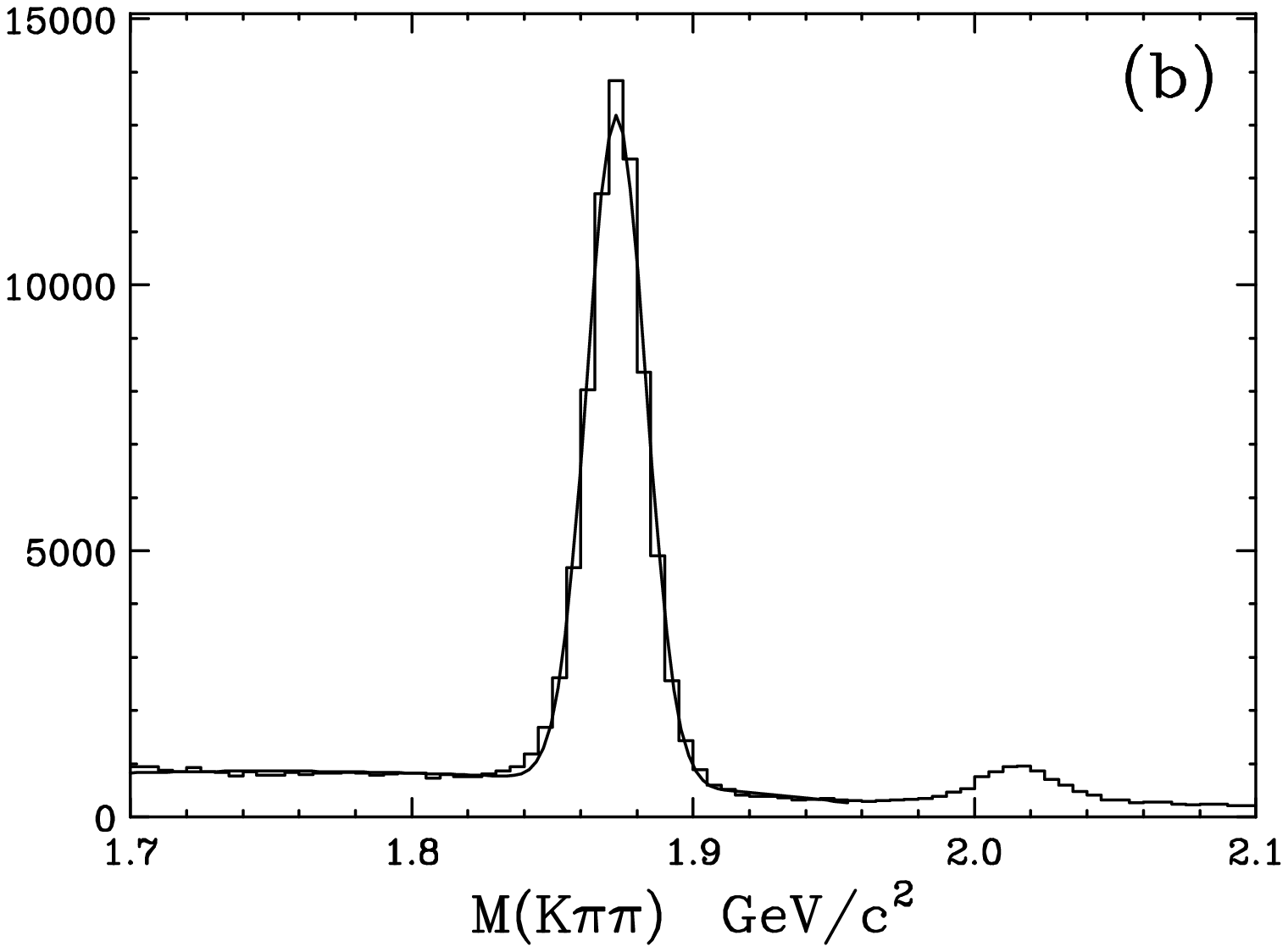}
\caption{We show two fitted \kpipi{} signals and fits (a) The $K^-
\pi^+ \pi^+$ mass spectra subjected to the baseline cuts.  The yield
obtained from this fit is $D^+$ yield of 116\,748 events.  (b) The
$K^- \pi^+ \pi^+$ mass spectra with an additional out-of-material cut
and secondary isolation cut. The yield
obtained from this fit is $D^+$ yield of 65\,421  events.
The peak near 2.01 \gevcsq{} is from the decay $D^{*+} \rightarrow D^0 \underline{\pi}^+
\rightarrow (K^- \pi^+ )\underline{\pi}^+$ where the $\underline{\pi}$ pion is included in the 
primary vertex thus mimicking the \kpipi{} decay topology.
\label{dplus}}
\end{center}
\end{figure}
The \kpipi{} yield was determined by fitting the $K^- \pi^+ \pi^+$
mass distributions to a Gaussian peak over a polynomial
background. Figure \ref{dplus} shows these fits for the baseline
sample and the sample with the additional out-of-material and
isolation cuts.  The \krzmnbrb{} branching ratio is derived by
dividing the fitted \krzb{} yield in \krzmndk{} events by the fitted
\kpipi{} peak yield in data and then comparing to that obtained in our
Monte Carlo with a known input branching ratio.  Because our
\kpimndk{} simulation model has an s-wave amplitude interfering with
\krzmndk{}, we correct our ratio by a factor of 0.945, the fraction of
Monte Carlo events due to the \krzmndk{} process alone. We
compute the numerator of this fraction, by integrating over the $K^-
\pi^+ \mu^+ \nu$ phase space the model intensity discussed in
Reference \cite{anomaly} with the s-wave amplitude set to zero. We
then divide this numerator by the intensity integral where both the
s-wave amplitude and \krzb{} amplitudes are set to their proper values
according to Reference \cite{anomaly}.  This approach is frequently
used in charm Dalitz plot analyzes to assess the fractional
contributions of various quasi-two-body decay channels such as $D_s^+
\rightarrow \phi \pi^+$ even though these channels should be described
by their quantum mechanical amplitudes rather than partial decay
rates.  After this correction, we obtain \krzmnbrb{} =
\krzresultstat{} where the quoted error is statistical only based on
our samples with baseline, out-of-material and isolation cuts.

Three basic approaches were used to determine the systematic error on
\krzmnbrb{}. In the first approach, we measured the stability of the
branching ratio with respect to variations in analysis cuts designed
to suppress backgrounds. In these studies we varied the secondary
isolation cut, the detachment cut, and a cut on the number of tracks
in our primary vertex. The square root of the sample variance of the
branching ratio from 8 such cut sets was 0.71 times our statistical
error. In the second approach, we split our sample according to a
variety of cuts applied to both the \kpimndk{} numerator and \kpipi{}
denominator and estimated a systematic based on the consistency of the
branching ratio among the split samples. We split our sample based on
the reconstructed $D^+$ momentum, particle versus antiparticle, and
the sum of the energy of $D^+$ secondary tracks that would strike our
calorimeter and thus contribute to our hadronic trigger. We chose
these variables since they significantly influence our acceptance.
For example the $D^+$ momentum is tied to both our geometrical
acceptance, and particle identification efficiencies. The calorimetric
energy for a fixed $D^+$ momentum is significantly lower for the
\kpimndk{} than the \kpipi{} final state. The maximum systematic error
for these splits was 1.57 times our statistical error. In the third
approach we checked the stability of the branching fraction as we
varied specific parameters in our Monte Carlo model and fitting
procedure.  These included varying the level of the background Monte
Carlo prior to subtracting, the power of the momentum dependence of
the width in the Breit-Wigner line shape as discussed previously, 
and the value of the three form factor ratios that
describe the \krzmndk{} decay distribution.  We estimate a combined
systematic from line shape, background level, and form factor
systematic that is also 1.57 times our statistical error. Combining
all three systematic error estimates in quadrature we have:
$$\krzmnbrb{} = \krzresult{}$$
We believe that our systematic error estimate is conservative.

\mysection{Analysis of the \phimndk{} final state}

\begin{figure}[tbph!]
 \begin{center}
\includegraphics[height=1.9in]{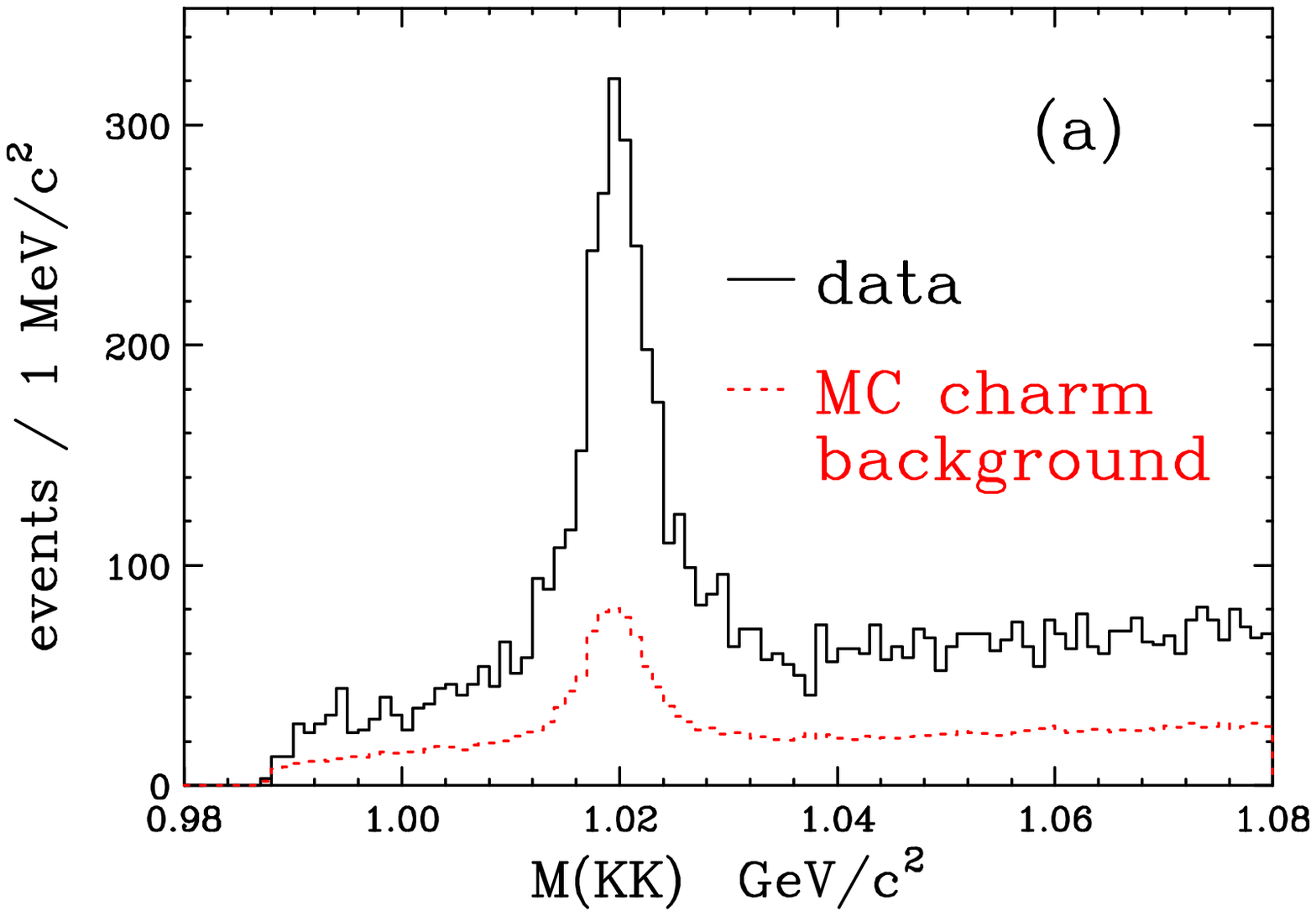}
\includegraphics[height=1.9in]{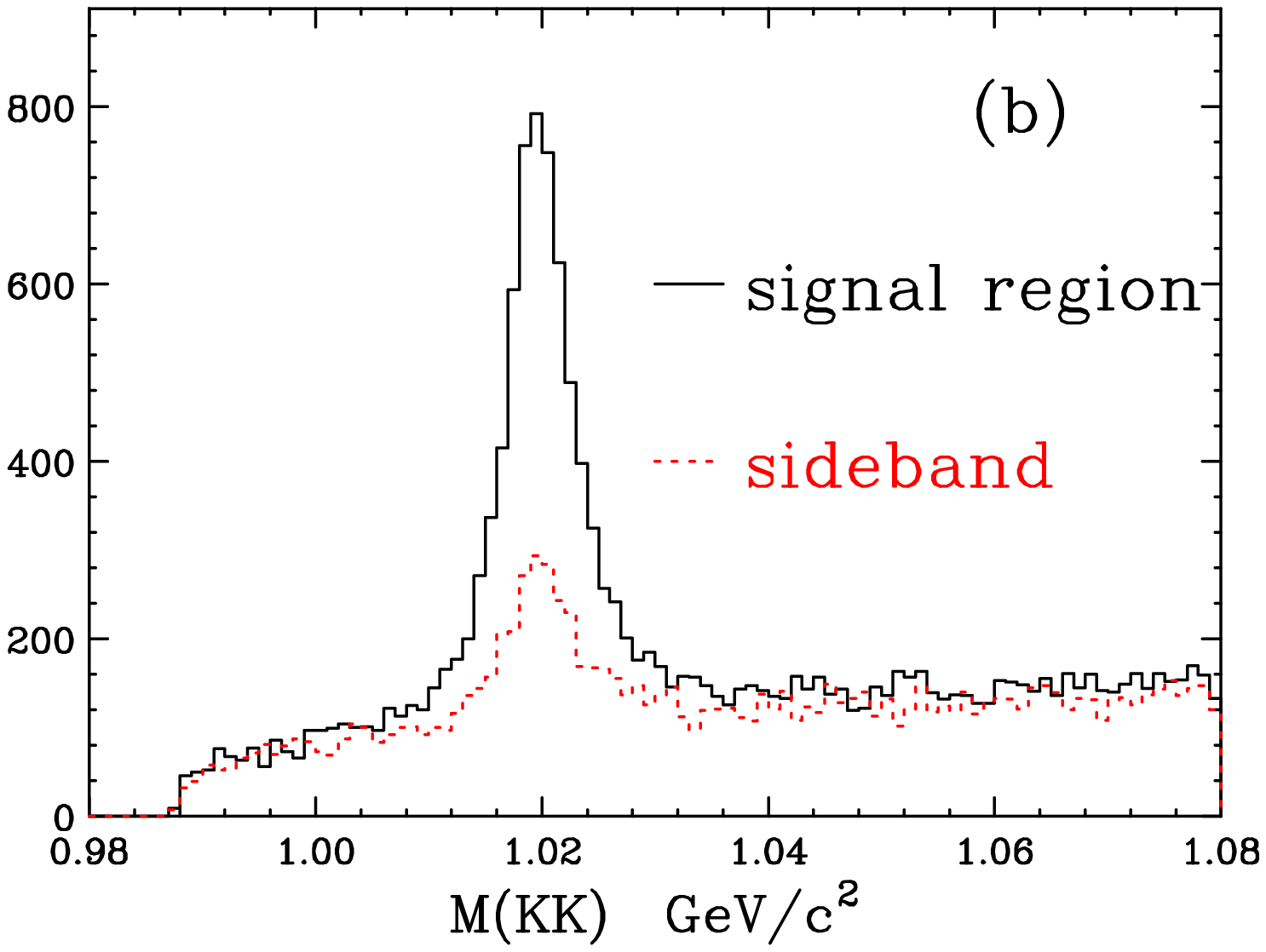}
\includegraphics[height=1.9in]{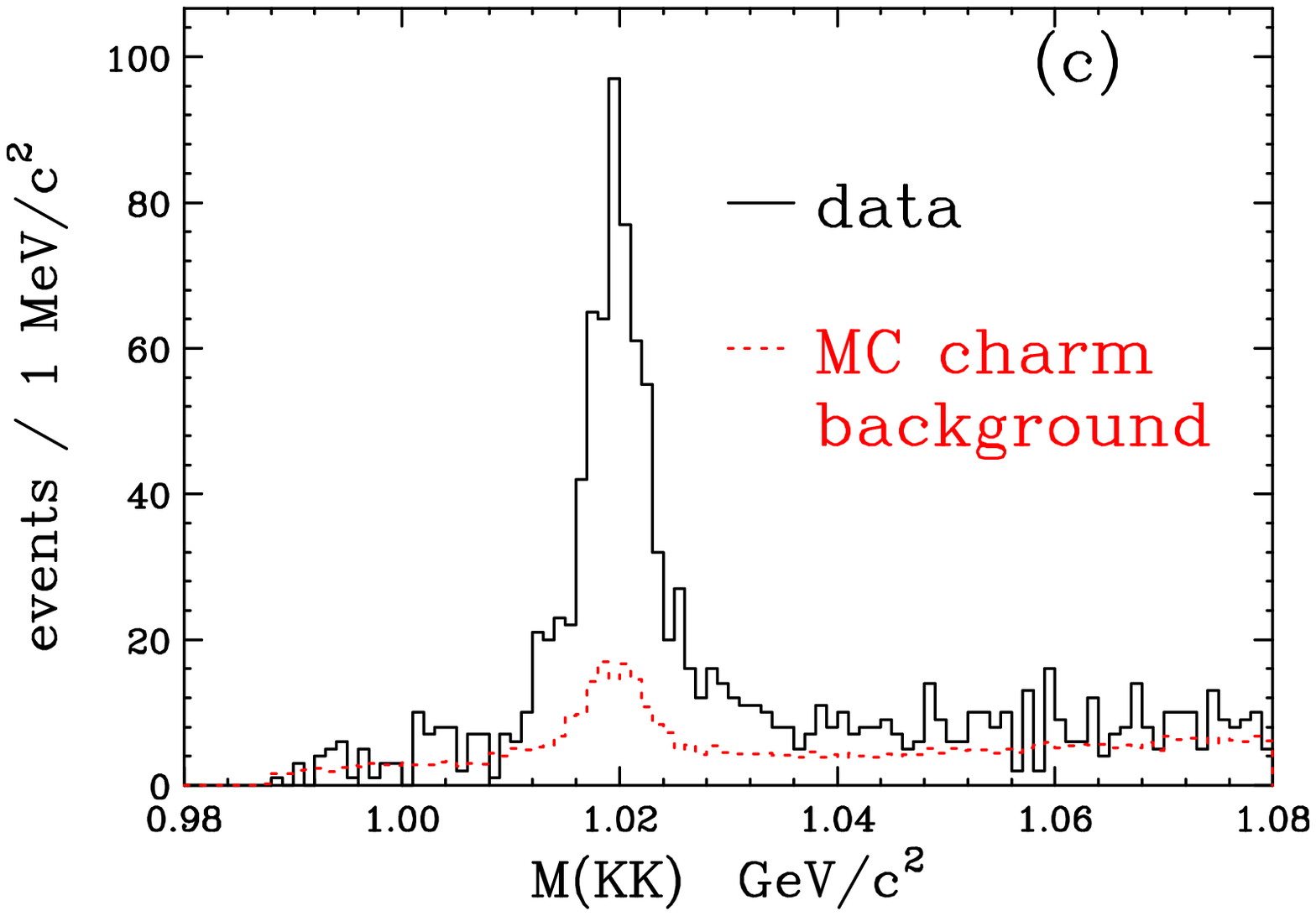}
\includegraphics[height=1.9in]{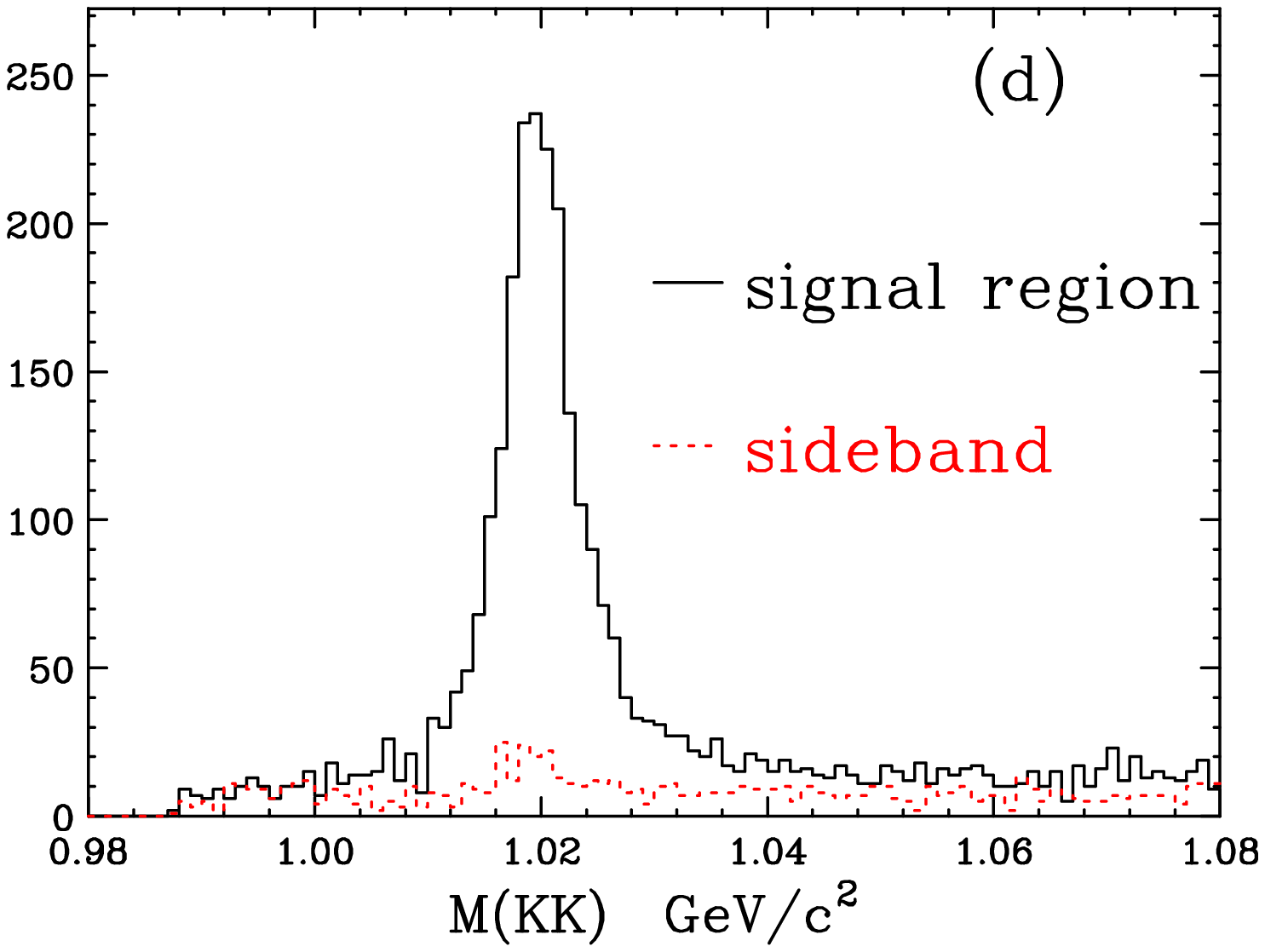}
\caption{We show signals for \phimndk{} and \phipi{} (a) The $K^- K^+$
mass spectra in \phimndk{} events subjected to the baseline cuts. The
data is plotted in a solid line and the background Monte Carlo is plotted in
a dotted line.  The difference in the fitted $\phi$ yield in the data and that
in the background Monte Carlo is 2682 events.  (b) The $K^- K^+$ mass
spectra in \phipi{} events subjected to the baseline cuts. The spectra
for the $D_s^+$ signal region
($1.946~\gevcsq{}<m_{KK\pi}<1.996~\gevcsq{}$) is plotted in a solid line and the
sideband region ($1.913~\gevcsq{} <m_{KK\pi}<1.938~\gevcsq{}$ and
$2.004~\gevcsq{} <m_{KK\pi}< 2.029~\gevcsq{}$) is plotted in a dotted line.  The
difference between the fitted $\phi$ yields in the signal and sideband
region is 4695 events. (c) The $K^- K^+$ mass spectra in \phimndk{}
events subjected to the baseline, out-of-material,and isolation cuts.
The background subtracted $\phi$ yield is 793 events.  (d) The $K^-
K^+$ mass spectra in \phipi{} events subjected to the baseline,
out-of-material, and isolation cuts. The sideband subtracted $\phi$
yield is 2192 events.
\label{phi_signal}}
\end{center}
\end{figure}

Figure~\ref{phi_signal} shows the $K^- K^+$ mass distributions that we
obtained using two versions of the cuts described above.  Figures
\ref{phi_signal} (a) and (c) compare the $K^+ K^-$ mass spectra for
\phimndk{} candidates and in a charm background Monte Carlo normalized
to the observed number of \phipi{} events where the \phimndk{}
contribution is excluded. The use of the out-of-material and isolation
cuts both significantly reduces the $\phi$ background level and
increases the agreement between the observed and that predicted
background from charm sources according to our Monte Carlo. The
difference of the fitted $\phi$ yields between the data and background
Monte Carlo is the yield of $\phi$ that we will attribute to
\phimndk{}.

Figures \ref{phi_signal} (b) and (d) compare the $K^+ K^-$ mass
spectra for \phipi{} candidates in a $D_s^+$ signal region and a
sideband region based on the $K^+ K^- \pi^+$ invariant mass.
The difference in fitted $\phi$ yields between the signal and sideband
$K^+ K^-$ mass spectra will form our estimate of the \phipi{} yield.

 The \phimnbrb{} branching ratio was derived by dividing the background
subtracted, fitted $\phi$ yield in \phimndk{} events by the 
sideband subtracted, fitted $\phi$ yield in data and then comparing to
the ratio obtained in our Monte Carlo with a known input branching
ratio. We obtained \phimnbrb{} = \phiresultstat{} where the quoted error
is statistical only based on our samples with baseline,
out-of-material and isolation cuts.

The systematic error on the \phimnbrb{} branching fraction was determined
in a way similar to that used for the \krzmndk{}. The systematic error estimate
obtained by varying analysis cuts was 1.06 times the statistical error. 
The systematic error estimate by splitting samples was  1.12 times
the statistical error. The final estimate due to varying fit parameters
such as the charm background level was 0.94 times the statistical error.
Combining all three sources we have:
$$\phimnbrb{}=\phiresult$$

\mysection{Summary}
\begin{table}[htp]
\caption{Measurements of the \krzlnbrb{} branching fraction}
\begin{center}
\begin{tabular}{l|l|l}
Group & electron & muon \\
\hline \hline
This work & & $0.602~\pm~0.010~\pm~0.021$ \\
CLEO \cite{cleo} & $0.74~\pm~0.04~\pm~0.05$ &  $0.72~\pm~0.10~\pm~0.06$\\
CLEO \cite{cleo_old} & $0.67~\pm~0.09~\pm~0.07$ & \\
E687 \cite{e687} & & $0.56~\pm~0.04~\pm~0.06$ \\
OMEGA \cite{omega} & $0.62~\pm~0.15~\pm~0.09$ & \\
ARGUS \cite{argus} & $0.55~\pm~0.08~\pm~0.10$ & \\
E653 \cite{e653} & & $0.46~\pm~0.07~\pm~0.08$ \\
E691 \cite{e691}   & $0.49~\pm~0.04~\pm~0.05 $ \\
\end{tabular}
\end{center}
\label{krztable}
\end{table}

Table \ref{krztable} summarizes measurements of the \krzlnbrb{}
branching fraction for electrons and muons.  Our measurement is the
first one to include the effects on the acceptance due to changes in
the decay angular distribution brought about by the s-wave
interference \cite{anomaly}.  
After correcting the muon numbers by a factor of 1.05
to compare to electrons according to the prescription of Reference
\cite{PDG}, we find that all values in the table are consistent with
their weighted average ($0.62 \pm 0.02$) with a confidence level of
19\% if systematic errors are added in quadrature with statistical
errors. Our number is about 1.57 standard deviations below the recent
CLEO measurement and about 2.1 standard deviations above the number
obtained by E691 \cite{e691}.

\begin{table}[htp]
\caption{Measurements of the \philnbrb{} branching fraction}
\begin{center}
\begin{tabular}{l|l|l}
Group & electron & muon \\
\hline \hline
This work & & $0.54~\pm~0.033~\pm~0.048$ \\
CLEO2 \cite{cleo2phi} & $0.54~\pm~0.05~\pm~0.04$ &  \\
E687 \cite{e687phi} & & $0.58~\pm~0.17~\pm~0.07$ \\
ARGUS \cite{argus} & $0.57~\pm~0.15~\pm~0.15$ & \\
CLEO \cite{cleophi} & $0.49~\pm~0.10~\pm~0.12$ & \\
\end{tabular}
\end{center}
\label{phitable}
\end{table}
Table \ref{phitable} summarizes measurements of the \philnbrb{}
branching fraction.  All results are remarkably consistent with an
average of 0.54 $\pm$ 0.04.

\mysection{Acknowledgments}

We wish to acknowledge the assistance of the staffs of Fermi National
Accelerator Laboratory, the INFN of Italy, and the physics departments
of the collaborating institutions. This research was supported in part
by the U.~S.  National Science Foundation, the U.~S. Department of
Energy, the Italian Istituto Nazionale di Fisica Nucleare and
Ministero dell'Universit\`a e della Ricerca Scientifica e Tecnologica,
the Brazilian Conselho Nacional de Desenvolvimento Cient\'{\i}fico e
Tecnol\'ogico, CONACyT-M\'exico, the Korean Ministry of Education, and
the Korean Science and Engineering Foundation.


\begin{thebibliography}{15}
\bibitem{anomaly} FOCUS Collab. (J.M. Link \etal),
        Phys. Lett.~B 535 (2002) 43.
\bibitem{KS} J.G.~Korner and G.A.~Schuler, 
        Z.~Phys.~C 46 (1990) 93.
\bibitem{cleo} CLEO Collab., {\it Measurement of the \krzlndk{}
        Branching Fraction}, Mar 20, 2002, hep-ex/0203030.
\bibitem{isgw2} D. Scora and N. Isgur, 
        Phys. Rev. D 62 (1995) 2783.
\bibitem{ycp} See for example, J.~M.~Link \etal, Phys. Lett.~B 485 (2000)
        62-70, and references therein. 
\bibitem{nim} E687 Collab., P.~L.~Frabetti \etal,
        Nucl.~Instrum.~Meth.~A 320 (1992) 519.
\bibitem{CNIM} FOCUS Collab., J.~M.~Link \etal, 
        Nucl.~Instrum.~Meth.~A 484 (2002) 270.
\bibitem{PDG} Particle Data Group, J.~Bartels \etal, 
        Eur.~Phys.~J.~C \textbf{15} (2000) 1.
\bibitem{cleo_old} CLEO Collab., A.~Bean \etal, 
        Phys. Lett.~B 317 (1993) 647.
\bibitem{e687} E687 Collab., P.~L.~Frabetti \etal,
        Phys. Lett.~B 307 (1993) 262.
\bibitem{omega} CERN WA82 Collab., M.~Adamovich \etal, 
        Phys. Lett.~B 268 (1991) 142.
\bibitem{argus} Argus Collab., H. Albrecht \etal, 
        Phys. Lett. B 255 (1991) 634.
\bibitem{e653} E653 Collab., K.~Kodama \etal, 
        Phys. Lett. B 286 (1992) 187.
\bibitem{e691} E691 Collab., J.~C.~Anjos \etal, 
        Phys. Rev. Lett. 62 (1989) 62.
\bibitem{cleo2phi} CLEO Collab., F.~Butler \etal, 
        Phys. Lett. B 324 (1994) 255.
\bibitem{e687phi} E687 Collab., P.~L.~Frabetti \etal,
        Phys. Lett. B 313 (1993) 253.
\bibitem{cleophi} CLEO Collab., J.~Alexander \etal, 
        Phys. Rev. Lett. 65 (1990) 1531.
\end{thebibliography}
\end{document}